\documentclass[12pt]{iopart}
\usepackage{iopams} 
\usepackage{graphicx}
\usepackage{xcolor}
\begin{document}

\title[]{Active microrheology to determine viscoelastic parameters of Stokes-Oldroyd B fluids using optical tweezers}

\author{Shuvojit Paul}

\address{Indian Institute of Science Education and Research, Kolkata}

\author{Avijit Kundu}

\address{Indian Institute of Science Education and Research, Kolkata}

\author{Ayan Banerjee}

\address{Indian Institute of Science Education and Research, Kolkata}
\ead{ayan@iiserkol.ac.in}

\begin{abstract}
We use active microrheology to determine the frequency dependent moduli of a linear viscoelastic fluid in terms of the polymer time constant ($\lambda$), and the polymer ($\mu_p$) and solvent viscosity ($\mu_s$), respectively. We measure these parameters from the response function of an optically trapped Brownian probe in the fluid under an external perturbation, and at different dilutions of the viscoelastic component in the fluid. This is an improvement over bulk microrheology measurements in viscoelastic Stokes-Oldroyd B fluids which determine  the complex elastic modulus $G(\omega)$ of the fluid, but do not, however, reveal the characteristics of the polymer chains and the Newtonian solvent of the complex fluid individually. In a recent work [Paul \textit{et al}., 2018 J. Phys. Condens. Matter \textbf{30} 345101], we linearized the Stokes-Oldroyd B fluid model and thereby explicitly formulated the frequency dependent moduli in terms of ($\mu_p$) and ($\mu_s$), which we now extend to account for an external sinusoidal force applied to the probe particle. We measure $\lambda$, $\mu_p$, and $\mu_s$ experimentally, and compare with existing the $\lambda$ values in the literature for the same fluid at some of the dilution levels, and obtain good agreement. Further, we use these parameters to calculate the complex elastic modulus of the fluid again at certain dilutions and verify successfully with existing data. This establishes our method as an alternate approach in the active microrheology of complex fluids which should reveal information about the composition of such fluids in significantly greater detail and high signal to noise.

\end{abstract}

\maketitle

\section{Introduction}
The fundamental difference between liquids and solids is their response under applied shear strain - while solids store energy and thus are elastic,  liquids dissipate energy and are therefore viscous in nature. However, fluids ranging from cytoplasm to ketchup, store and dissipate mechanical energy in relative proportions depending on frequency. Therefore, they are called viscoelastic. There exists a strong interest in the scientific community to understand and measure the parameters of viscoelastic fluids mainly because, the biological entities which sustain life are viscoelastic \cite{ahmed2018active, turlier2016equilibrium,ayala2016rheological,prado2015viscoelastic,brust2013rheology,larson1999structure,ferry1980viscoelastic}. The rheological properties of such fluids are often parameterized in terms of a frequency dependent complex elastic modulus $G^{*}(\omega)$ whose real part $G'(\omega)$ remains in phase with the applied strain and represents the storage of energy (elastic part), while the imaginary part $G''(\omega)$ remains out of phase, and represents the loss (viscous part) of energy in the system \cite{ferry1980viscoelastic, mason1995optical, doi1988theory, Tassieri}. The complex dynamic viscosity is given by $\eta (\omega)=G^{*}(\omega)/(-i\omega)$. Typically, the bulk rheological properties of a viscoelastic material is measured by analyzing its response when the entire sample is subject to an external strain. Therefore, the local heterogeneity in the sample remains unexamined \cite{crocker2000two, neckernuss2015active, weigand2017active}. Additionally, this method commonly requires $\sim$ mL of samples which may limit its use for expensive or scarce samples, such as biological fluids. The invention of optical tweezers in 1986 by Ashkin and colleagues \cite{ashkin1986observation}, has facilitated 'microrheology' (rheology in the micrometer scale) with $\sim\mu$L of samples and the above-mentioned issues have been overcome \cite{brau2007passive, waigh2016advances}. In this method, typically, the Brownian motion (passive microrheology) or the motion under external perturbation (active microrheology) of a micron sized trapped particle inside a fluid is studied to extract the frequency dependent viscoelastic parameters $G'(\omega)$ and $G''(\omega)$. Active microrheology understandably provides enhanced capabilities and wider parameter space of rheological measurements along with better signal-to-noise ratio (S/N) over the passive technique \cite{robertson2018optical, paul2018two}.

To get deeper insight into the sample property, different models describing a linear viscoelastic fluid have been developed. Foremost among these, is the Maxwell model \cite{gotze2008complex, boon1991molecular} which has been further developed into the generalized Maxwell model or Jeffreys' model \cite{raikher2013brownian, raikher2010theory}. The high degree of simplification \cite{volkov1996non, volkov1990theory, raikher1996dynamic} used in the Maxwell model ease out calculations, but the model sometimes fails to interpret experimental results. It has been shown that at least the Jeffrey's model is required to explain and understand experimental results in detail \cite{wilhelm2009magnetic}. Both of these models can provide the stress and shear-strain relation for a linear viscoelastic fluid in terms of the fundamental parameters of the fluid. For example, the Maxwell model describes a time constant $\tau_{M}$ which marks a transition from the high frequency elastic nature of the sample to the low frequency viscous regime, while the Jeffrey's model contains a zero-frequency viscosity $\eta_{0}$ and a correction term as the background viscosity $\eta_{\infty}$ \cite{grimm2011brownian}. However, these models are based on the bulk properties of the viscoelastic fluid and do not provide any information about its basic constituents. To address this issue, we have shown in a recent work that a viscoelastic fluid can be understood as a viscous solvent which contains a polymer network mixed with it. We have demonstrated that the background viscosity $\eta_{\infty}$ is nothing but the solvent contribution to the zero-frequency viscosity, while $\eta_{0}$ is the polymer contribution to it, while the Maxwell time constant is basically the polymer time constant \cite{paul2018free}. We obtained this understanding by linearizing the Stokes-Oldroyd B equations for small perturbations and for low Weissenberg number. Clearly, this approach links the overall rheological behavior of the fluid with the characteristics of its constituents and provides greater acuity in measurement and understanding of viscoelasticity itself. 

In this paper, we measure for the first time, the polymer and solvent contributions to the viscoelasticity of a linear viscoelastic fluid having a single time constant. Thus, we experimentally determine the phase response of a micron sized spherical particle confined in a harmonic potential in such a Stokes-Oldryod B fluid under an external perturbation. First, we solve the equation of motion of a trapped particle under external perturbation in a fluid as described in the recent work \cite{paul2018free}. Then, we fit the expression of the phase response to  experimentally measured data to extract the parameter values.  Typically, in the passive microrheology technique, a generalized Stokes-Einstein relation is employed to convert the mean-squared displacement (MSD) of the probe into the complex elastic modulus $G^{*}(\omega)$. This process involves a fourier transform of the MSD, which is rather non-trivial in practice, given a finite set of data points over a finite time domain \cite{Tassieri}. Understandably, viscoelastic fluids with very low concentration of the polymer network, can be very easily analyzed by this simple method. In addition, this method involves measuring the MSD which is obtained from the amplitude of Brownian motion of a probe particle. However, we have shown recently that phase measurement using a lock-in is more sensitive and accurate than amplitude measurement \cite{paul2018two}, which is not unexpected since the amplitude of a signal gets more effected by noises than the phase. Further, the measurement from the phase does not require the conversion of the signal into real displacement units, so that errors involving in determining the calibration factor can be avoided (which, incidentally, is significantly affected by detector electronics). For consistency check,  we have applied this phase-measurement based technique to normal water and obtained good agreement in our determination of the solvent viscosity. Further, we have proceeded to measure the viscoelastic parameters for samples of Polyacrylamide (PAM) to water solutions at different dilution levels. We observe that our results are in good agreement with that reported recently \cite{chandra_shankar_das_2018}, which have been performed  for relatively low polymer concentration solutions. For  solutions of higher polymer concentrations, however, the measured polymer contributions to the viscosities are not satisfactory. We believe this to be due to the inherent ineffectiveness of our model in dealing with the non-linear nature of viscoelasticity or the additional complexity resulting in the superposition of several time constants and other parameters that the high concentration of polymer would induce in a fluid \cite{mason1995optical}.  For linear viscoelastic fluids and for low-concentrations, our work opens a new approach in  microrheology and can be used very extensively due to its simple methodology and ease-of-use.
\section{Theory}
\label{Theory}
The equation of motion describing the trajectory of a spherical particle of mass $m$ confined in a harmonic potential of force constant $k$ in a linear viscoelastic fluid in the Cartesian co-ordinate system (we choose $x$ here) is given by 
\begin{equation}
m\ddot{x}(t)=-\int_{-\infty}^{t}\gamma (t-t')\dot{x}(t')dt' -k[x(t) - x_{0}(t)]+\xi (t)
\label{eq1}
\end{equation}
where, the integral term on the right-hand side incorporates a generalized time-dependent memory kernel, and $\gamma (t)$, represents damping by the fluid, so that it can be termed as the time dependent friction coefficient. $x(t)$ and $x_{0} (t)$ are the instantaneous positions of the particle and the potential minimum, respectively, and $\xi (t)$ is the Gaussian-distributed correlated thermal noise due to the random collisions of the fluid molecules with the particle which leads to the Brownian motion of the particle. The correlation of the noise is given by $\langle \xi (t)\xi(t')\rangle=2k_{B}T\gamma (t-t')$ where $k_{B}$ is the Bolzmann constant and $T$ is the temperature. Due to the negligible mass of the trapped  probe particle and the fact that we are at low Raynold's number, the momentum relaxation time scale and the vorticity time scale are negligible compared to typical experimental time scales. Thus, we neglect the inertial term from Eq.~\ref{eq1} and average over the noise since we are basically interested in the response function of the particle under external perturbation. Therefore, the equation of motion in the frequency domain can be effectively written as
\begin{equation}
-i\omega \gamma(\omega)x(\omega)+k x(\omega)=k x_{0}(\omega)
\label{eq2}
\end{equation}
Now, $\gamma (\omega)$ of a Stokes-Oldroyd B fluid is related to the polymer time constant $\lambda$, the polymer and the solvent contribution to the viscosity $\mu_{p}$ and $\mu_{s}$, respectively, as \cite{paul2018free}
\begin{equation}
\gamma(\omega)=6\pi\mu_{s} a_{0}\left(1+\frac{\mu_{r}}{-i\omega\lambda+1}\right)
\label{eq3}
\end{equation}
where $a_{0}$ is the radius of the trapped particle and $\mu_{r}=\mu_{p}/\mu_{s}$. Substituting Eq.~\ref{eq3} in Eq.~\ref{eq2} and calculating the phase of the response of the particle we get
\begin{equation}
\Phi\left(\omega\right)=\tan^{-1} \left[\frac{\frac{1+\mu_{r}}{\lambda^{2}}\omega + \omega^{3}}{\frac{k}{\lambda^{2}\gamma_{0}} + \left(\frac{k}{\gamma_{0}} + \frac{\mu_{r}}{\lambda}\right)\omega^{2}}\right]
\label{eq4}
\end{equation}
where $\gamma_{0}=6\pi \mu_{s}a_{0}$. Therefore, it is possible to fit the experimentally measured phase with Eq.~\ref{eq4} to infer the characteristic parameters of the concerned fluid. Later on, these parameters can be employed to obtain the complex shear modulus of the fluid which is given by $G^{*}(\omega)=-i\omega \gamma(\omega)/6\pi a_{0}$ for a spherical probe particle \cite{paul2018free} .

\begin{figure}[h]
	\centering
	\includegraphics[scale=0.6]{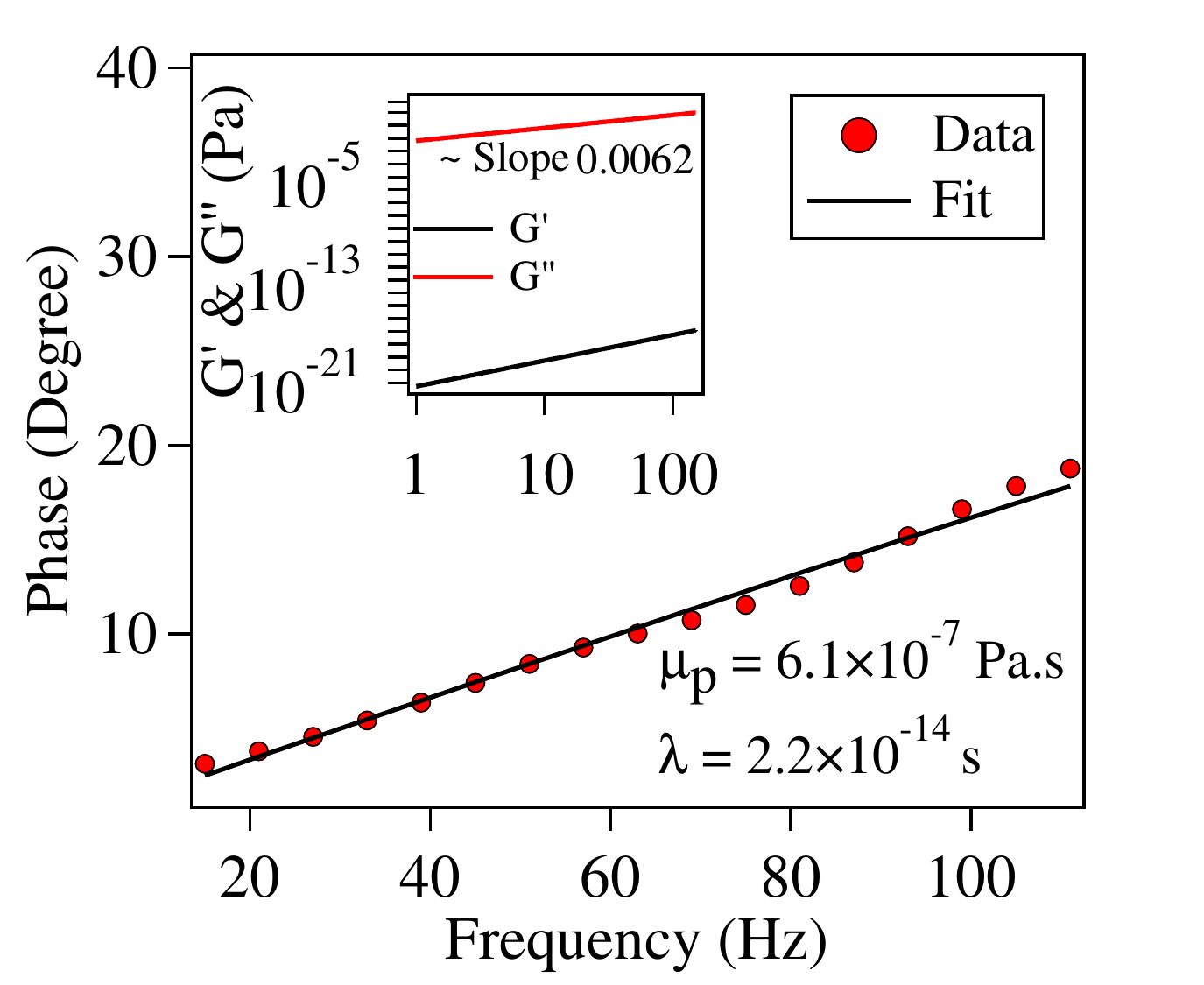}
	\caption{Phase response of the spherical probe of radius $1.5$ $\mu$m in water as a function of driving frequency along with a fit to the data using Eq.~\ref{eq4}. The measured trap stiffness is $52(4)$ $\mu$N/m. In the inset, $G'(f)$ and $G''(f)$ have been plotted against frequency - these have been calculated theoretically using the extracted parameter values of the fluid. The loss part $G''(f)$ increases linearly with frequency having a slop of $2\pi\mu_{s}=0.0062$. This is the case for a purely viscous fluid. Understandably, $G''(f)$ is close to zero. $\lambda$ and $\mu_{p}$ also tend to zero which imply the effective phase response to be $\phi (f)=\tan^{-1}\left(2\pi\gamma_{0}f/k\right)$. This is also is valid for a viscous fluid.}
	\label{fig1}
\end{figure}

\begin{figure}[h]
	\centering
	\includegraphics[scale=0.6]{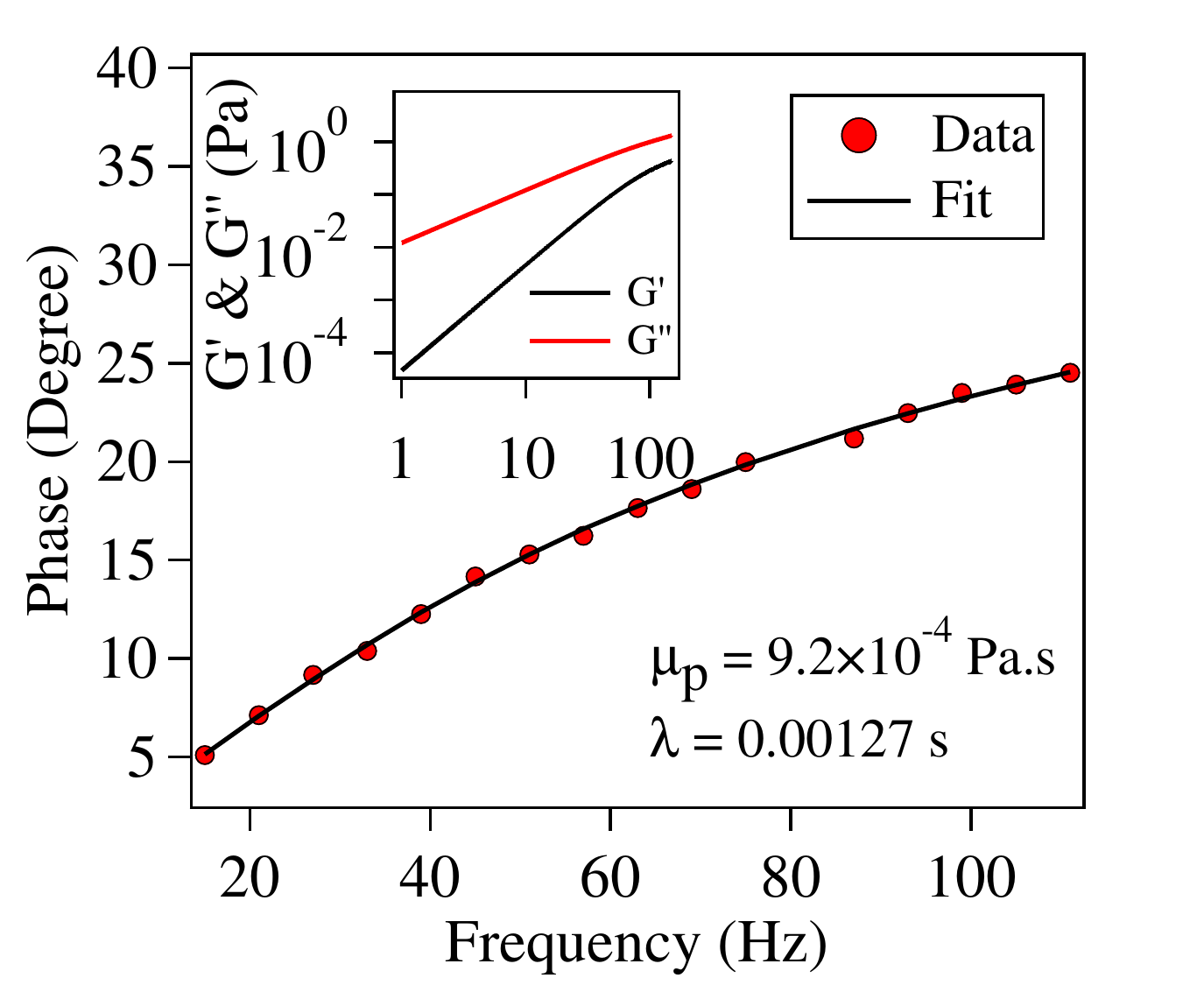}
	\caption{Phase response of the probe particle of radius $1.5$ $\mu$m in 0.008$\%$ w/w PAM to water solution with driving frequency along with a fit to the data using Eq.~\ref{eq4}. The measured trap stiffness is $48(3)$ $\mu$N/m. In the inset, $G'(f)$ and $G''(f)$ have been plotted against frequency which has been calculated theoretically using the extracted parameter values of the fluid.}
	\label{fig2}
\end{figure}

\section{Experimental Details}
\label{Experiment}
We perform the experiments using an optical tweezers built around an inverted microscope (Zeiss Axiovert.A1 Observer) with an objective lens (Zeiss PlanApo 100x, 1.4 numerical aperture) tightly focusing a laser beam of wavelength 1064 nm into the sample. For detecting the displacement of trapped particles, we employ a co-propagating laser of wavelength 780 nm. A balanced detection system \cite{paul2017, paul2018two, Bera2017} placed at the back-focal plane of the objective detects the back-scattered light from the trapped probe particle to track its position. We modulate the trapping laser beam at different frequencies by a piezo-mirror placed at the conjugate plane of the objective focal plane and keep the detection beam fixed. Simultaneously, we use a CCD camera to image the trapped particle. Our sample is inserted into a sealed sample chamber which is prepared by attaching a glass slide to a cover slip by double-sided tape so that the dimensions become around $20 \times 10 \times 0.2$ mm. For the viscoelastic sample, we have taken water-based polyacrylamide  (PAM, flexible polyelectrolytes, $M_{w}=(5-6)\times10^{6} $ gm/mol, Sigma-Aldrich) and trapped single  spherical polystyrene probe particles of radius $1.5$ $\mu m$ around 30 $\mu m$ away from the nearest wall to get rid of surface effects. We modulate the trap center sinusoidally by the piezo-mirror with an amplitude of 110 nm at different frequencies and record the response for 60 seconds at each frequency by a data acquisition card (NI USB-6356). Simultaneously, the data is fed into a lock-in amplifier (Standford Research, SR830) and averaged over the same time duration to measure the relative phase of the response of the particle with respect to the modulation. In the absence of the modulation, we have recorded the Brownian motion of the particle to calculate the trap stiffness.    

\section{Results and discussions}
\label{analysis}
\begin{figure}[h]
	\centering
	\includegraphics[scale=0.55]{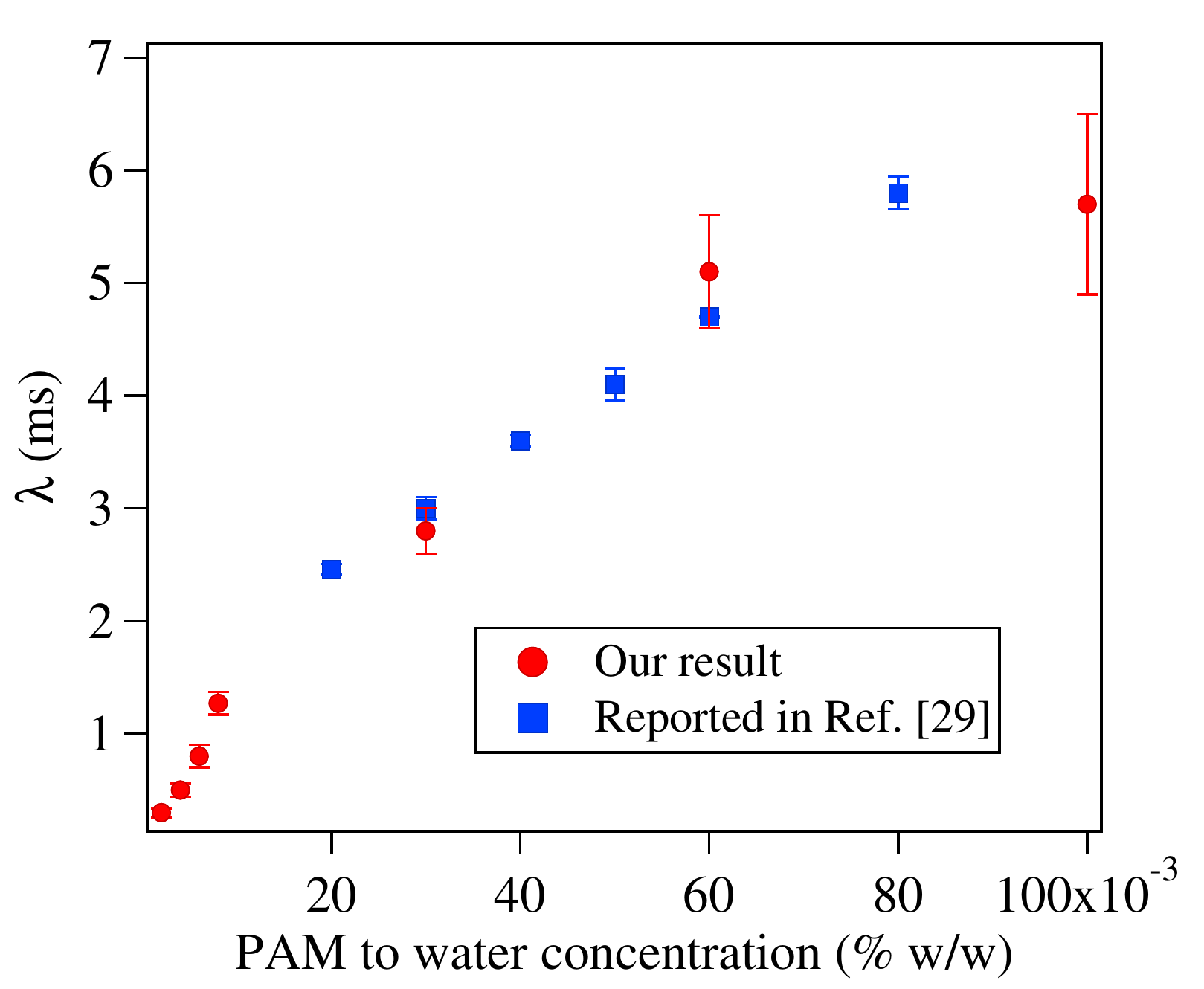}
	\caption{Measurement of the polymer time constant $\lambda$ as a function of PAM to water concentration using our technique superimposed with the values reported in the literature \cite{chandra_shankar_das_2018}. The trends of the variation of $\lambda$ with concentration for our values and those reported earlier clearly appear to be similar. }
	\label{fig3}
\end{figure}

In equilibrium, the stiffness of the optical trap can be measured using the equipartition theorem since the latter is independent of the rheological property of the sample. According to this theorem,  the trap stiffness in our system is given by $k=k_{B}T/\langle\left(x - \langle x \rangle\right)^{2}\rangle$. After we determine the trap stiffness, we determine the fluid parameters for different concentrations by plotting the measured phase of the probe response as a function of driving frequency, and fitting the data to Eq.~\ref{eq4}. The fit parameters then yield the parameters of the fluid. To check for the efficacy of our technique, we have first performed the measurement for pure water and obtained good agreement. The results are demonstrated in Fig.~\ref{fig1} along with the corresponding storage ($G'$) and loss ($G''$) moduli in the inset. The time constant $\lambda$ and the polymer contribution of viscosity $\mu_{p}$  are both close to zero, and the solvent contribution of viscosity is $0.0009\pm0.0001$ Pa.s, which is very close to the viscosity of water at 300K. Note that this effectively changes Eq.~\ref{eq4} to $\phi (\omega)=\tan^{-1}\left(\gamma_{0}\omega/k\right)$, which is indeed the phase response of a particle trapped in a pure viscous fluid. The storage modulus is almost zero whereas the loss modulus increases with frequency having slope $0.0062$ which is equal to $2\pi\times\mu_{s}$, \cite{2018arXiv180804796P} thus corresponding to a purely viscous sample. In Fig. \ref{fig2}, we show a typical phase response of a linear viscoelastic fluid ($0.008\%$ w/w PAM to water solution) with driving frequency and the fit along with the corresponding storage and loss moduli in the inset. Clearly, in the comparison with that of water, ($G'$) is much greater ($\sim 10^{17}$). It is important to point out that for both the cases we kept the trap stiffness fixed. We then evaluated the parameter values for PAM to water solutions of different concentrations, which are shown in Table~\ref{tb1}. The evaluated solvent contribution of viscosity is $\mu_{s} = 0.0009 \pm 0.0002$ Pa.s from all the measurements. The time constant increases with the concentration of the solution and almost saturates after $0.1\%$. Our measurements of $\lambda$ are consistent with a recent work (Fig.~5 of Ref.~\cite{chandra_shankar_das_2018}) - for ease of comparison, we have juxtaposed their data with ours in Fig. \ref{fig3}. Clearly, our measurements follow the same trend in the variation of $\lambda$ against frequency as reported in this paper. On the other hand, the polymer contribution of viscosity also increases with the PAM concentration in the solution  as expected. The small deviations of our measurements from the reported values can be due to the differences in experimental conditions including local temperature, the molecular weight of PAM, electronic noise, etc. Furthermore, our measurements are really localised involving a very small region of the fluid, whereas the reported measurements are bulk in nature, so that differences may appear due to local temperature fluctuations, density variations due to inhomogeneous mixing, and other local effects. Further, it is clear from  Table~\ref{tb1} that the measured polymer contribution to the viscosity $\mu_{p}$ does not appear reliable for higher concentrations, with the change of $\mu_{p}$ being rather small with large change of concentration. This basically suggests that at these levels of viscoelasticity, where the nature of the viscoelastic response may become non-linear to applied strain, or additional time constants may appear \cite{mason1995optical}, our theory may have limitations since it only accounts for linear viscoelastic response and a single time constant. It is thus likely that we are measuring a superposition of different time constants at higher polymer concentrations, and thereby obtaining erroneous results.
\begin{table}[tb]
	\centering
	\small
	\caption{\ Extracted parameters with varying PAM concentrations in water. The stiffness of the trap has been kept fixed at $k=48(3)$ $\mu$N/m over the measurements. The extracted solvent viscosity from all the measurements is $\mu_{s}=0.0009(2)$ Pa.s.}
	\label{tb1}
	\vspace{0.01\textheight}
	\begin{tabular*}{0.6\textwidth}{@{\extracolsep{\fill}}lll}
		\hline
		concentration ($\%$ w/w) & $\lambda$ (s) & $\mu_{p}$ (Pa.s) \\
		\hline
		0.002 & 0.00031(4) & 0.00036(5) \\
		0.004 & 0.00053(6) & 0.00032(4) \\
		0.006 & 0.0008(1) & 0.00047(7) \\
		0.008 & 0.0013(1) & 0.00083(4) \\
		0.03 & 0.0028(2) & 0.0012(1)\\
		0.06 & 0.0051(5) & 0.0015(1)\\
		0.1  & 0.0057(8) & 0.0014(2)\\
		0.5  & 0.0033(5) & 0.0015(2)\\
		1   & 0.006(1)  & 0.004(1)\\
		\hline
	\end{tabular*}
\end{table}

\section{Conclusions}
\label{con}
In conclusion, we have presented a simple experimental method  to extract the rheological parameter values of a linear viscoelastic fluid using optical tweezers. Our method employs active microrheology, which straightaway enhances the signal to noise of the measurements. Thus, we measure the phase response of a Brownian probe particle that we modulate sinusoidally in an optical trap at different frequencies. The inherent construction of our approach - based on linearizing the Stokes-Oldroyd B equation for viscoelastic fluids - provides for a more profound understanding about the constituents of such fluids inasmuch that it reveals the polymer and solvent characteristics separately. Our method has a basic advantage over the most commonly used technique of characterizing viscoelastic fluids from measurements of the storage and loss parameters $G'(\omega)$ and $G''(\omega)$ which involve a complex discrete fourier transformation of a finite set of MSD data over a finite time, which we are able to avoid entirely. This Fourier transform can be erroneous for the low fluid concentrations that are required in microrheology. Furthermore, our approach of measuring the phase using a lock-in amplifier has an obvious signal to noise advantage over techniques which measure the amplitude of motion of Brownian particles and are therefore much more susceptible to experimental noise. The phase measurement also precludes the requirement of the calibration of the particle displacement in real physical units for which a careful measurement of the detector sensitivity is essential, which naturally leads to enhanced systematic errors. We test our technique on a purely viscous fluid - water, for which we obtain very good agreement with well-known literature values, and different viscoelastic solutions of PAM and water where the concentration of the former is varied. At low polymer concentrations, we obtain rather reliable measurements of the time constant $\lambda$ which match with values in literature \cite{chandra_shankar_das_2018}, while for increased polymer concentrations, the values seem to be unreliable with very small change in the measured $\mu_{p}$ with increasing concentration. This we attribute to the limitations of our theory in the case of non-linear viscoelastic fluids and for more complex fluids with additional time constants, which probably is the case when we increase the concentration of PAM in the solution. We also calculate $G'(\omega)$ and $G''(\omega)$ for the different fluid concentrations, and obtain expected trends against frequency. This is an entirely new approach in microrheology, and we intend to extend our measurements to more diverse systems  which still fit our model in their viscoelastic response such as blood or plasma (and biological fluids, in general - they being weakly viscoelastic) - where the accuracy of the technique may also render it as a useful diagnostic tool by comparing the viscoelastic parameters in infected and normal conditions. We are presently commencing these experiments.
\section{Acknowledgments}
This work was supported by IISER Kolkata, an autonomous teaching and
research institute supported by the Ministry of Human Resource Development,
Govt. of India.

\section*{References}
\bibliographystyle{iopart-num}
\bibliography{VE_parameter}
\end{document}